\documentclass[pra,twocolumn,superscriptaddress,10pt,noshowpacs,floatfix]{revtex4}
\usepackage[english]{babel}
\usepackage[T1]{fontenc}
\usepackage[utf8]{inputenc}
\usepackage{graphicx,epstopdf}
\usepackage{amssymb}
\usepackage{amsmath}

\DeclareMathOperator{\arcsinh}{arcsinh}
\usepackage{amsfonts}
\usepackage{bbm}
\usepackage{color}
\usepackage{latexsym}
\usepackage{caption}
\usepackage{subcaption}
\usepackage{float}
\usepackage{times,txfonts}
\usepackage{color,soul}
\usepackage{listings}
\usepackage{bbold}
\usepackage{xcolor}

\newcommand{\beq}{\begin{equation}}
\newcommand{\eeq}{\end{equation}}
\newcommand{\bea}{\begin{eqnarray}}
\newcommand{\eea}{\end{eqnarray}}
\newcommand{\orcidEdson}{}
\newcommand{\orcidJonathan}{}
\newcommand{\orcidmca}{}
\newcommand{\orcidfurtado}{}
\begin{document}

\title{White Dwarf Stellar Structure from Effective Polymer Geometry in Loop Quantum Gravity}

\author{Edson Otoniel\orcidEdson\!\!}
\email{edson.otoniel@ufca.edu.br}
\affiliation{Universidade Federal do Cariri (UFCA), Instituto de Forma\c{c}\~ao de Educadores - IFE, R. Oleg\'ario Emidio de Araujo S/N, Brejo Santo - CE, 63.260-000 - Brazil}

\author{Jonathan A. Rebou\c{c}as\orcidJonathan\!\!}
\email{jalvesreboucas@ifce.edu.br}
\affiliation{Instituto Federal de Educa\c{c}\~ao, Ci\^encia e Tecnologia do Cear\'a (IFCE), Iguatu, Brazil}

\author{M. C. Ara\'{u}jo\orcidmca}
\email{michelangelo.araujo@ufca.edu.br}
\affiliation{Universidade Federal do Cariri, Av. Tenente Raimundo Rocha, \\
Cidade Universit\'{a}ria, Juazeiro do Norte, Cear\'{a}, 63048-080, Brazil}

\author{J. Furtado\orcidfurtado}
\email{job.furtado@ufca.edu.br}
\affiliation{Universidade Federal do Cariri, Av. Tenente Raimundo Rocha, \\
Cidade Universit\'{a}ria, Juazeiro do Norte, Cear\'{a}, 63048-080, Brazil}

\begin{abstract}

We construct an effective Tolman Oppenheimer Volkoff system for cold carbon white dwarfs using the areal radius form of a polymer metric sector motivated by loop quantum gravity. The two asymptotic mass parameters of the geometry are retained in the stellar prescription through $M_B\rightarrow m(R)$ and $M_W=\eta m(R)$, while the polymer amplitude is controlled by $A_\lambda$. The matter sector is kept fixed and is described by the Chandrasekhar equation of state and by the same carbon model with the Coulomb lattice correction. The resulting equations recover the general relativistic TOV system and the symmetric polymer limit. For the undeformed sequences we obtain $M_{\max}=1.4166\,M_\odot$ for the Chandrasekhar model and $M_{\max}=1.3850\,M_\odot$ when the lattice correction is included. Turning on $A_\lambda$ shifts the massive part of the equilibrium sequence upward without stiffening the equation of state, reaching $M_{\max}=1.7125\,M_\odot$ and $1.6907\,M_\odot$ at $A_\lambda=100$ for the two matter models. These configurations remain within the matter domain imposed by the inverse beta decay boundary used in the scan. The asymmetric ratio $\eta=M_W/M_B$ changes the metric function near the polymer transition region, but its effect on white dwarf observables is small: across the selected configurations, $M_{\max}$ changes by less than $0.1\%$ and the corresponding radius by less than $0.33\%$. The calculation therefore identifies $A_\lambda$ as the parameter controlling the super Chandrasekhar displacement of the mass radius relation, while $\eta$ acts mainly as a geometric asymmetry parameter in the low compactness regime probed by white dwarfs.
\end{abstract}

\maketitle


\section{Introduction}

White dwarfs are useful stellar systems in which gravity can be tested against a quantum pressure scale. Their equilibrium is governed mainly by electron degeneracy pressure, while the limiting mass follows from the relativistic behavior of the degenerate gas \cite{Chandrasekhar1931}. Corrections to this ideal picture are also well identified. The Coulomb contribution of the lattice changes the pressure and energy density of cold matter \cite{Salpeter1961,HamadaSalpeter1961}, and inverse beta decay sets a composition dependent upper density domain for massive configurations \cite{Chamel2013}. These ingredients make white dwarfs useful reference objects: the matter model is comparatively simple, but the maximum mass and the radius of dense configurations remain sensitive to the gravitational structure equations.

The relativistic stellar problem is normally formulated through the Tolman Oppenheimer Volkoff equations, obtained from static spherical solutions of Einstein's equations coupled to a perfect fluid \cite{Tolman1939,OppenheimerVolkoff1939}. Although white dwarfs are less compact than neutron stars, relativistic corrections are not irrelevant near the Chandrasekhar scale. General relativistic calculations of massive white dwarfs show that radii and surface quantities can differ from Newtonian estimates in the high mass regime \cite{Rotondo2011,Carvalho2018}. For this reason, the general relativistic sequence is the appropriate baseline before introducing any deformation of the geometry.

Massive and overluminous type Ia supernovae have also kept the question of super Chandrasekhar progenitors open. The event SNLS 03D3bb was interpreted as requiring a progenitor above the usual Chandrasekhar scale \cite{Howell2006}, and strongly magnetized white dwarf models were later proposed as one possible route to larger masses \cite{DasMukhopadhyay2013,Otoniel2019Magnetized}. Such interpretations are restricted by reaction thresholds and stability filters, including electron capture and pycnonuclear processes \cite{Chamel2013,Malheiro2021Nuclear}. In the present work this observational context is used only as motivation for controlled equilibrium calculations. We do not fit supernova data, and we do not replace a stability analysis by the existence of static solutions.

Compact stars are often used to test modified gravity because a change in the gravitational sector generally changes the hydrostatic balance and therefore the mass and radius of an equilibrium sequence. Reviews of stellar structure in modified theories emphasize that matter effects and gravity effects must be separated, since equation of state uncertainties can mimic changes in the structure equations \cite{Olmo2020}. White dwarfs are especially useful in this respect: their microphysics is simpler than that of neutron stars, and their mass radius relation has already been used to constrain modified gravity scenarios \cite{Jain2016,Carvalho2017ModifiedWD,Otoniel2026RTLmWD}. Related quantum gravity phenomenology in white dwarfs has often entered through changes in the uncertainty relation or in the degenerate equation of state \cite{Wang2012,Rashidi2016}. Our approach follows a different route by keeping the carbon equation of state fixed and changing the effective metric sector.

The gravitational sector considered in this work is inspired by effective polymer models developed within loop quantum gravity (LQG). Although general relativity (GR) provides an exceptionally successful description of gravitation and accounts for phenomena such as gravitational-wave emission \cite{LIGOScientific:2016aoc}, gravitational lensing \cite{Fomalont:2009zg}, and accretion around compact objects \cite{Abramowicz:2011xu}, the singularity theorems show that Einstein’s field equations generically lead to spacetime singularities under physically reasonable conditions \cite{Penrose:1964wq,Hawking:1970zqf}. The resulting breakdown of the classical description at sufficiently large curvatures has motivated several approaches to quantum gravity, including LQG \cite{Bojowald:2007ky,Bojowald:2001xe,Ashtekar:2011ni,Olmedo:2017lvt}, string theory \cite{Cornalba:2002fi,Berkooz:2002je}, and noncommutative geometry \cite{Maceda:2003xr,Gorji:2014pka}. Among these proposals, LQG offers a background-independent and nonperturbative framework in which quantum-geometric effects can be incorporated through effective polymerized spacetime geometries. In polymer black-hole models, the Schwarzschild singularity is replaced by a spacelike transition surface joining black-hole and white-hole regions, while the classical Schwarzschild exterior is recovered sufficiently far from the high-curvature domain \cite{Ashtekar:2020ifw,Bodendorfer:2019cyv,Bodendorfer:2019nvy}. The magnitude of the quantum corrections is controlled by the polymer parameter $A_{\lambda}$, and the two asymptotic regions may be characterized by the black-hole and white-hole masses $M_B$ and $M_W$, respectively. This class of effective metrics has already been employed to investigate rotating extensions and the orbital dynamics of test particles \cite{Brahma:2020eos,Tu:2023xab}, while effective LQG dynamics coupled to perfect fluids has also been shown to modify the equations governing static stellar equilibrium, leading to Planck-scale star configurations \cite{WilsonEwing2025}. These developments suggest that polymer corrections need not be restricted to black-hole interiors and motivate the central question addressed here: how does the equilibrium structure of an ordinary compact star change when the matter equation of state is kept fixed, but the classical gravitational sector is replaced by an effective polymer inspired metric?

In this paper we construct an effective white dwarf stellar system from the areal radius form of the polymer geometry. The two mass parameters of the metric are kept distinct through the prescription $M_B\rightarrow m(R)$ and $M_W=\eta m(R)$, while $A_\lambda$ controls the polymer deformation. We derive the corresponding effective TOV equations, recover the general relativistic limit, and use two cold carbon equations of state: the Chandrasekhar model and the same model with the lattice correction. The numerical part then examines how the equilibrium sequences respond to $A_\lambda$ and to the asymmetric ratio $\eta=M_W/M_B$. This separates the shift in the mass radius relation from the metric asymmetry itself, and fixes the domain in which the configurations remain below the inverse beta decay boundary used in the equation of state.

The paper is organized as follows. Section~\ref{sec:polymer_geometry} rewrites the polymer metric in areal radius and defines the stellar prescription for the two mass parameters. Section~\ref{sec:effective_tov} derives the effective structure equations and their general relativistic limit. Section~\ref{sec:eos} specifies the cold carbon equations of state and the inverse beta decay boundary. Section~\ref{sec:results} presents the mass radius sequences, the selected maximum mass configurations, and the metric diagnostic for the parameter $\eta$. The final section summarizes the physical interpretation and the limits of the present calculation.

\section{Polymer geometry in areal radius}\label{sec:polymer_geometry}

The stellar problem requires the polymer geometry to be written in variables with direct meaning for equilibrium configurations. We therefore begin with the exterior metric sector and then recast its radial part in terms of the areal radius.

We start from the effective polymer metric used in Refs.~\cite{Bodendorfer:2019nvy,Bodendorfer:2019cyv,Brahma:2020eos,Tu:2023xab}. In coordinates adapted to the polymer solution, the line element is
\begin{equation}
ds^2
=
-8A_{\lambda}M_B^2\mathcal{A}(r)dt^2
+
\frac{dr^2}{8A_{\lambda}M_B^2\mathcal{A}(r)}
+
\mathcal{B}(r)d\Omega^2,
\label{eq:polymer_metric_r}
\end{equation}
with
\begin{align}
\mathcal{A}(r)
&=
\frac{1}{\mathcal{B}(r)}
\left(1+\frac{r^2}{8A_{\lambda}M_B^2}\right)
\left(1-\frac{2M_B}{\sqrt{8A_{\lambda}M_B^2+r^2}}\right),
\label{eq:polymer_A_r}\\
\mathcal{B}(r)
&=
\frac{
512A_{\lambda}^{3}M_B^4M_W^2
+
\left(r+\sqrt{8A_{\lambda}M_B^2+r^2}\right)^6
}{
8\sqrt{8A_{\lambda}M_B^2+r^2}
\left(\sqrt{8A_{\lambda}M_B^2+r^2}+r\right)^3
}.
\label{eq:polymer_B_r}
\end{align}
The constants $M_B$ and $M_W$ label the two asymptotic Schwarzschild regions of the effective geometry. In the original vacuum solution, $A_\lambda$ is related to the polymer scale and to these asymptotic masses by $A_{\lambda}=\left[\lambda_k/(M_BM_W)\right]^{2/3}/2$. In the stellar construction below, however, $A_\lambda$ is treated as a fixed effective deformation parameter. The replacement $M_B\to m(R)$ is applied to the metric functions at fixed $A_\lambda$, rather than to the vacuum relation defining $A_\lambda$. In the black to white hole interpretation, the classical singularity is replaced by a spacelike transition surface located at the minimum of the areal radius. We use this metric sector as the starting point for the stellar ansatz below.

For a stellar calculation the radial coordinate must be the areal radius, since the surface radius and the stellar mass sequence are reported in that variable. We therefore set
\begin{equation}
R^2=\mathcal{B}(r),
\end{equation}
and introduce the auxiliary positive variable
\begin{equation}
y
=
\left[
\frac{r+\sqrt{8A_\lambda M_B^2+r^2}}
{\sqrt{8A_\lambda M_B^2}}
\right]^2.
\label{eq:y_definition}
\end{equation}
The asymmetric stellar prescription used here is
\begin{equation}
\begin{aligned}
M_B&\rightarrow m,
\qquad
M_W=\eta m,\\
A_\lambda&=\mathrm{constant},
\qquad
\eta=\mathrm{constant}.
\end{aligned}
\label{eq:mass_ansatz_parameter}
\end{equation}
At this stage $m$ is treated as a parameter; it is promoted to an interior mass function in the next section.

With Eq.~(\ref{eq:mass_ansatz_parameter}), the areal radius takes the implicit form
\begin{equation}
R^2=2A_\lambda m^2 h(y,\eta),
\qquad
h(y,\eta)=\frac{y^3+\eta^2}{y(y+1)}.
\label{eq:areal_h}
\end{equation}
Equivalently, the selected value of $y$ is determined by the cubic equation
\begin{equation}
2A_\lambda m^2y^3
-R^2y^2
-R^2y
+2A_\lambda\eta^2m^2
=0.
\label{eq:y_cubic}
\end{equation}
In the numerical implementation we use the positive root that approaches
\begin{equation}
y\sim \frac{R^2}{2A_\lambda m^2}
\qquad
(A_\lambda\rightarrow0^+),
\label{eq:y_gr_root}
\end{equation}
so that the exterior Schwarzschild form is recovered in the GR limit.

The radial coefficient in areal radius is not simply the time component of the metric. The transformation from $r$ to $R$ gives
\begin{equation}
\frac{dr^2}{8A_\lambda m^2\mathcal{A}(r)}
=
\frac{dR^2}{F},
\qquad
F
=
8A_\lambda m^2\mathcal{A}(r)
\left(\frac{dR}{dr}\right)^2.
\end{equation}
In terms of the implicit variable $y$, this radial function can be written as
\begin{equation}
F(y,A_\lambda,\eta)
=
Q(y,A_\lambda)y^2
\left(\frac{h_y}{h}\right)^2,
\label{eq:F_def}
\end{equation}
where
\begin{equation}
h_y
=
\frac{\partial h}{\partial y}
=
\frac{y^4+2y^3-2\eta^2y-\eta^2}{y^2(y+1)^2},
\end{equation}
and
\begin{equation}
Q(y,A_\lambda)
=
1-\sqrt{\frac{2}{A_\lambda}}\frac{\sqrt{y}}{y+1}.
\label{eq:Q_def}
\end{equation}
The areal radius metric function used below is therefore the composition
\begin{equation}
F(R,m;A_\lambda,\eta)
=
F\!\left[y(R,m;A_\lambda,\eta),A_\lambda,\eta\right].
\end{equation}

Two limits fix the normalization of the construction. For $\eta=1$ one obtains
\begin{equation}
h(y,1)=\frac{y^3+1}{y(y+1)}
=
y-1+\frac{1}{y},
\end{equation}
and the areal radius relation reduces to the symmetric polymer expression used in the earlier derivation. On the positive root defined by Eq.~(\ref{eq:y_gr_root}), the GR limit gives
\begin{equation}
\lim_{A_\lambda\to0^+}F(R,m;A_\lambda,\eta)
=
1-\frac{2m}{R}.
\label{eq:F_GR_limit}
\end{equation}
These two limits fix the normalization used in the stellar equations.

\section{Effective Tolman Oppenheimer Volkoff system}\label{sec:effective_tov}

The areal radius form obtained above can now be used as an interior stellar ansatz. The matter sector is kept as an isotropic perfect fluid, while the polymer correction enters through the radial metric function.

We promote the mass parameter in the areal radius metric to an interior mass function,
\begin{equation}
M_B\rightarrow m(R),
\qquad
M_W=\eta m(R).
\label{eq:mass_ansatz_function}
\end{equation}
The gravitational mass reported for a stellar configuration is
\begin{equation}
M_\star=m(R_\star),
\end{equation}
where $R_\star$ is the surface radius. This is not a baryonic mass definition. In particular, neither $M_B$ nor $M_W$ is identified with a rest mass observable in the present calculation.

The interior metric is written in areal radius as
\begin{equation}
ds^2
=
-e^{\nu(R)}dt^2
+
\frac{dR^2}{F(R,m;A_\lambda,\eta)}
+
R^2d\Omega^2.
\label{eq:effective_metric}
\end{equation}
The matter source is an isotropic perfect fluid,
\begin{equation}
T^\mu{}_\nu
=
\mathrm{diag}(-\rho,P,P,P),
\label{eq:perfect_fluid}
\end{equation}
closed by an equation of state $\rho=\rho(P)$. The matter prescription used to close the system is given in Sec.~\ref{sec:eos}. Thus the LQG input enters through the metric sector, while the matter model remains the standard white dwarf fluid used in the numerical sequences.

For the metric in Eq.~(\ref{eq:effective_metric}), the mixed Einstein tensor components needed for the stellar equations are
\begin{equation}
G^t{}_t
=
\frac{RF'(R)+F(R)-1}{R^2},
\label{eq:Gtt_areal}
\end{equation}
and
\begin{equation}
G^R{}_R
=
\frac{F(R)-1}{R^2}
+
\frac{F(R)\nu'(R)}{R}.
\label{eq:Grr_areal}
\end{equation}
The angular component is not introduced as an independent tangential pressure in this version. The evolved variables are only $m(R)$ and $P(R)$.

Since $F$ depends on $R$ both explicitly through the areal radius and implicitly through $m(R)$, its total radial derivative is
\begin{equation}
F'(R)=F_R+F_m\frac{dm}{dR},
\label{eq:F_total_derivative}
\end{equation}
where
\begin{equation}
F_R=\left(\frac{\partial F}{\partial R}\right)_{m,A_\lambda,\eta},
\qquad
F_m=\left(\frac{\partial F}{\partial m}\right)_{R,A_\lambda,\eta}.
\end{equation}
These partial derivatives are evaluated through the implicit root $y$. Differentiating Eq.~(\ref{eq:y_cubic}) gives
\begin{equation}
C_y
=
6A_\lambda m^2y^2-R^2(2y+1),
\label{eq:Cy_def}
\end{equation}
and therefore
\begin{equation}
y_R
=
\frac{2Ry(y+1)}{C_y},
\qquad
y_m
=
-\frac{4A_\lambda m(y^3+\eta^2)}{C_y}.
\label{eq:y_derivatives}
\end{equation}
Because Eq.~(\ref{eq:F_def}) has no further explicit dependence on $R$ or $m$ once $y$ is specified, we use
\begin{equation}
F_R=F_y y_R,
\qquad
F_m=F_y y_m.
\label{eq:F_derivatives}
\end{equation}
The denominators $C_y$, $F$, and $F_m$ are monitored in the numerical scan.

The mass equation follows from $G^t{}_t=-8\pi\rho$. Substituting Eq.~(\ref{eq:F_total_derivative}) into Eq.~(\ref{eq:Gtt_areal}) gives
\begin{equation}
R\left(F_R+F_m\frac{dm}{dR}\right)
+F-1
=
-8\pi\rho R^2.
\end{equation}
Solving for the derivative of the mass function yields
\begin{equation}
\frac{dm}{dR}
=
\frac{1-F-RF_R-8\pi\rho R^2}{RF_m}.
\label{eq:asym_mass}
\end{equation}

The pressure equation follows from the radial field equation and energy momentum conservation. From $G^R{}_R=8\pi P$,
\begin{equation}
\nu'
=
\frac{8\pi PR^2+1-F}{RF}.
\end{equation}
For minimally coupled isotropic matter,
\begin{equation}
\nabla_\mu T^\mu{}_\nu=0
\end{equation}
gives $dP/dR=-(\rho+P)\nu'/2$. Hence
\begin{equation}
\frac{dP}{dR}
=
-(\rho+P)
\frac{8\pi PR^2+1-F}{2RF}.
\label{eq:asym_pressure}
\end{equation}

Equations~(\ref{eq:asym_mass}) and~(\ref{eq:asym_pressure}) form the effective stellar system solved in this work. In the limit $A_\lambda\to0^+$, Eq.~(\ref{eq:F_GR_limit}) gives
\begin{equation}
F\to1-\frac{2m}{R},
\qquad
F_R\to\frac{2m}{R^2},
\qquad
F_m\to-\frac{2}{R},
\end{equation}
and the standard GR equations are recovered:
\begin{equation}
\frac{dm}{dR}=4\pi R^2\rho,
\end{equation}
\begin{equation}
\frac{dP}{dR}
=
-\frac{(\rho+P)(m+4\pi R^3P)}{R(R-2m)}.
\end{equation}
For $\eta=1$, the system also reproduces the symmetric polymer benchmark obtained from the explicit areal radius derivation.

\section{Equation of state}\label{sec:eos}

The stellar sequences are computed for cold carbon matter with no rotation, magnetic pressure, or thermal correction. This choice keeps the matter sector fixed and separates the geometric deformation from effects that require a different stellar model, such as magnetic quantization, magnetic stresses, and rapid rotation \cite{Otoniel2015MagneticFermions,Otoniel2021FastSpinningWD}. The role of the equation of state is therefore to provide the thermodynamic relation between pressure and energy density in Eqs.~(\ref{eq:asym_mass}) and~(\ref{eq:asym_pressure}). In this section we use natural units, $c=\hbar=1$. The numerical tables are converted to the geometrized units of the stellar equations before integration.

The electron degeneracy is described by the dimensionless Fermi momentum
\begin{equation}
x=\frac{p_F}{m_e}
=\frac{(3\pi^2 n_e)^{1/3}}{m_e},
\label{eq:eos_fermi_momentum}
\end{equation}
with chemical potential
\begin{equation}
\mu_e=m_e\sqrt{1+x^2}.
\label{eq:eos_mu_e}
\end{equation}
For carbon, the electron density entering Eq.~(\ref{eq:eos_fermi_momentum}) is fixed by charge neutrality,
\begin{equation}
n_e=Y_e n_b,
\qquad
Y_e=\frac{Z}{A}=\frac{1}{2}.
\label{eq:eos_charge_neutrality}
\end{equation}

The ideal degenerate electron gas then gives the pressure and electron energy density \cite{Chandrasekhar1931},
\begin{align}
p_e&=
\frac{m_e^4}{8\pi^2}
\left[
x\left(\frac{2x^2}{3}-1\right)\sqrt{1+x^2}
+\arcsinh x
\right],
\label{eq:eos_electron_pressure}\\
\mathcal{E}_e&=
\frac{m_e^4}{8\pi^2}
\left[
x(1+2x^2)\sqrt{1+x^2}
-\arcsinh x
\right].
\label{eq:eos_electron_energy}
\end{align}
This is the Chandrasekhar equation of state used as the reference matter model. The second matter model adds the electrostatic contribution of a one component body centered cubic lattice. With the sign convention adopted here,
\begin{equation}
\mathcal{E}_L=C_L e^2 n_e^{4/3}Z^{2/3},
\qquad
p_L=\frac{\mathcal{E}_L}{3},
\qquad
C_L<0.
\label{eq:eos_lattice}
\end{equation}
The lattice term therefore reduces the pressure at fixed density while leaving the carbon composition unchanged. This is the correction used in the Hamada Salpeter type white dwarf equation of state \cite{Salpeter1961,HamadaSalpeter1961}.

The thermodynamic source is then
\begin{equation}
p=p_e
\quad
\mathrm{or}
\quad
p=p_e+p_L,
\label{eq:eos_total_pressure}
\end{equation}
depending on whether the lattice term is omitted or retained, and
\begin{equation}
\mathcal{E}
=
\mathcal{E}_{\rm rest}
+\mathcal{E}_e
\quad
\mathrm{or}
\quad
\mathcal{E}
=
\mathcal{E}_{\rm rest}
+\mathcal{E}_e
+\mathcal{E}_L.
\label{eq:eos_total_energy}
\end{equation}
After conversion to geometrized units, the quantity denoted by $\rho$ in Eq.~(\ref{eq:perfect_fluid}) is the energy density source associated with $\mathcal{E}$.

The upper density domain is fixed by inverse beta decay. We use the Gibbs free energy per baryon,
\begin{equation}
\mathcal{G}=\frac{\mathcal{E}+p}{n_b},
\label{eq:eos_gibbs}
\end{equation}
to state the condition for the capture channel
\begin{equation}
(A,Z)+\Delta Z\,e^-
\rightarrow
(A,Z-\Delta Z)+\Delta Z\,\nu_e.
\label{eq:eos_capture_channel}
\end{equation}
The corresponding atomic mass threshold is
\begin{equation}
\mu_e^\beta
=
\frac{
\left[
M_{\rm atom}(A,Z-\Delta Z)-M_{\rm atom}(A,Z)
\right]
}{\Delta Z}
+m_e.
\label{eq:eos_beta_threshold}
\end{equation}
The transition is obtained from equality of the parent and daughter Gibbs free energies at fixed pressure, with the lattice contribution included in the thermodynamic balance \cite{Chamel2013}. For carbon this gives the adopted boundary
\begin{equation}
\rho_\beta\simeq4.16\times10^{10}\,
\mathrm{g\,cm^{-3}},
\qquad
\mu_e\simeq14.18\,\mathrm{MeV}.
\label{eq:eos_beta_numbers}
\end{equation}
The numerical sequences are stopped before this boundary. The condition defines the validity domain of the equation of state and is not used as a dynamical collapse calculation. Related studies of massive white dwarfs show that electron capture, pycnonuclear reactions, composition, rotation, and magnetic fields can restrict the admissible part of an equilibrium sequence \cite{Otoniel2019Magnetized,Malheiro2021Nuclear,Otoniel2021FastSpinningWD}.

\section{Results}\label{sec:results}

We now evaluate the effective structure equations for cold carbon white dwarf matter. In this regime the general relativistic sequence provides a sharp reference scale: the maximum mass is set by the relativistic degenerate electron pressure, while the radius decreases rapidly in the dense part of the sequence. White dwarfs therefore probe the polymer deformation in a low compactness regime where geometric corrections are measured against a well established mass and radius scale. The comparison with the undeformed sequence fixes the physical size of the effect.

The two polymer parameters play different roles. The parameter $A_\lambda$ changes the equilibrium sequence itself, shifting the massive configurations and increasing the maximum mass within the regularity domain of the scan. The asymmetric ratio $\eta=M_W/M_B$, on the other hand, enters the metric sector more directly than it enters the global stellar observables. Its effect appears in the function $\mathcal{B}(r)$, but it produces only small changes in the calculated masses and radii for the configurations considered here. All masses reported in this section are gravitational masses, and the configurations should be read as equilibrium solutions; radial stability is not analyzed in the present version.

\begin{figure}[H]
\centering
\includegraphics[width=\columnwidth]{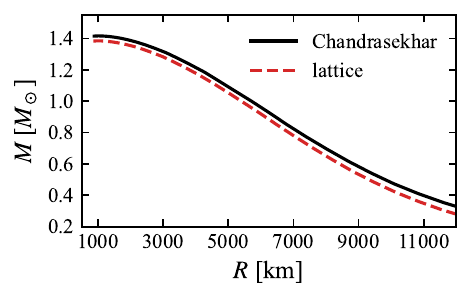}
\caption{General relativistic mass radius baseline for the two equations of state. The solid black curve corresponds to the Chandrasekhar equation of state, and the dashed red curve includes the lattice correction.}
\label{fig:gr_mr_baseline}
\end{figure}

Figure~\ref{fig:gr_mr_baseline} gives the undeformed reference sequence. The Chandrasekhar equation of state reaches $M_{\max}=1.4166\,M_\odot$ at $R=1010.7$ km, while the inclusion of the lattice correction lowers the maximum to $M_{\max}=1.3850\,M_\odot$ at $R=980.5$ km. This reduction is physically consistent with a small loss of pressure support at high central density. The change is not large enough to alter the qualitative Chandrasekhar behavior, but it sets the equation of state scale against which the polymer correction is compared.

\begin{figure}[H]
\centering
\includegraphics[width=\columnwidth]{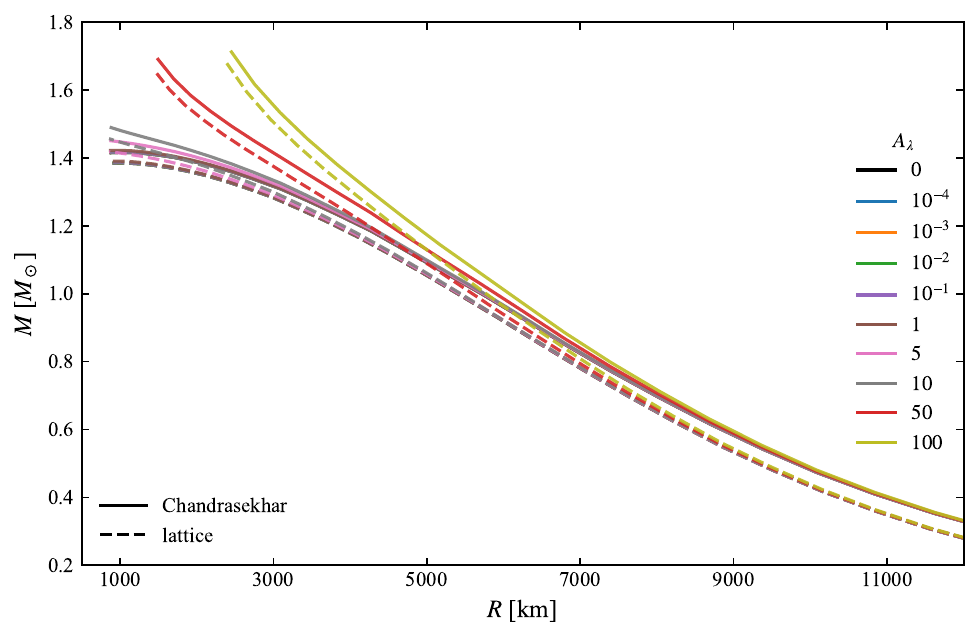}
\caption{Sequences of mass as a function of radius in the symmetric case $\eta=1$ for selected values of $A_\lambda$. Colors identify $A_\lambda$, while the line style distinguishes the two equations of state. Solid curves correspond to the Chandrasekhar equation of state and dashed curves to the equation of state with the lattice correction.}
\label{fig:lqg_mr_eta1}
\end{figure}

The effect of $A_\lambda$ is concentrated where the star is already close to the relativistic degeneracy limit. In Fig.~\ref{fig:lqg_mr_eta1}, small values of $A_\lambda$ leave most of the sequence close to the GR curve, while larger values move the most massive configurations to higher gravitational masses. In the effective TOV equations this behavior is not an equation of state stiffening; the matter model is held fixed. It is a geometric contribution to the balance between gravity and pressure gradients. The lattice sequence follows the same displacement, indicating that the polymer deformation acts on the structure equations rather than on a particular feature of the carbon equation of state.

\begin{figure}[H]
\centering
\includegraphics[width=\columnwidth]{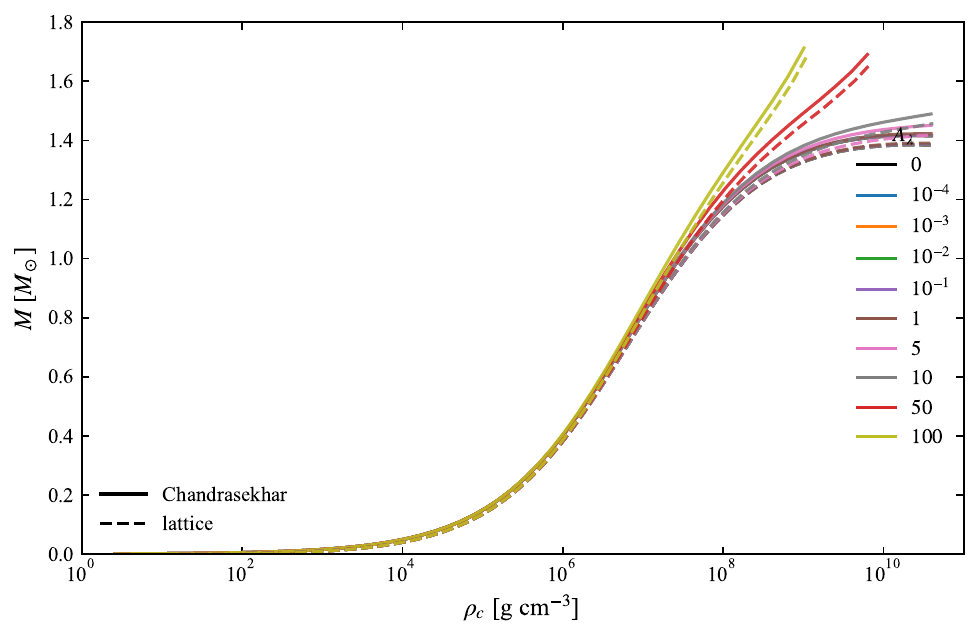}
\caption{Mass as a function of central mass density for the symmetric case $\eta=1$. The curves use the same parameter set and line convention as Fig.~\ref{fig:lqg_mr_eta1}. Colors identify $A_\lambda$, solid curves correspond to the Chandrasekhar equation of state, and dashed curves include the lattice correction.}
\label{fig:lqg_mrho_eta1}
\end{figure}

The same displacement is visible when the configurations are parametrized by the central density. Figure~\ref{fig:lqg_mrho_eta1} shows that the high mass part of the sequence moves upward as $A_\lambda$ grows, while the matter input remains the same. For the largest values of $A_\lambda$, the accepted points terminate at lower central densities because the metric domain filters remove part of the high density grid. This behavior is useful physically: the super Chandrasekhar masses reported below are not obtained by pushing the carbon equation of state beyond the inverse beta decay boundary.

\begin{figure}[H]
\centering
\includegraphics[width=\columnwidth]{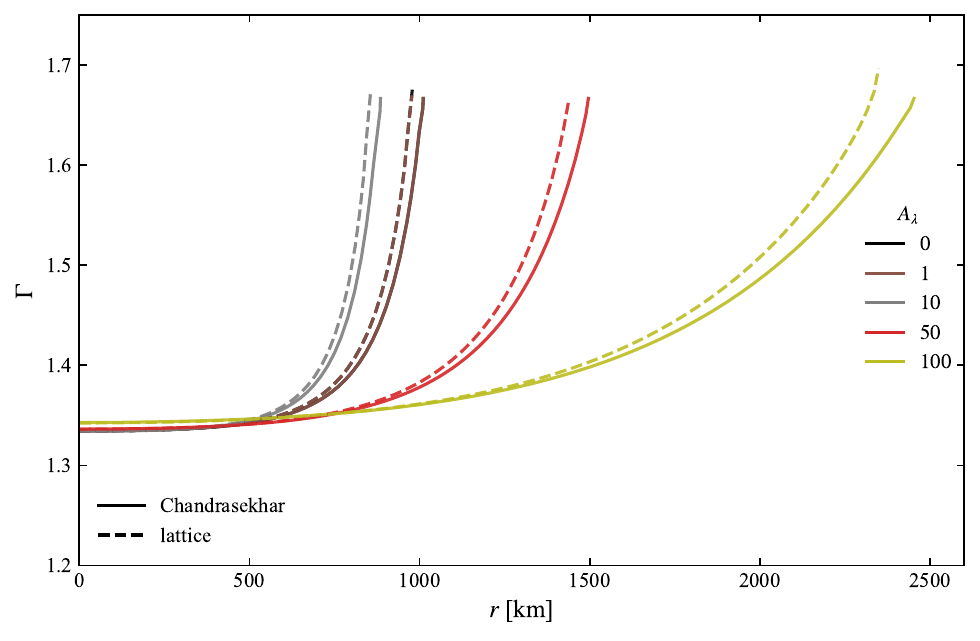}
\caption{Relativistic adiabatic index as a function of radius for representative maximum mass configurations in the symmetric case $\eta=1$. The displayed values are $A_\lambda=0,1,10,50,100$. Colors identify $A_\lambda$, while line styles distinguish the two equations of state as in Fig.~\ref{fig:lqg_mr_eta1}.}
\label{fig:lqg_gamma_eta1}
\end{figure}

The local thermodynamic response also supports this interpretation. We compute the relativistic adiabatic index,
\begin{equation}
\Gamma
=
\frac{\rho+P}{P}\frac{dP}{d\rho},
\label{eq:gamma_ad}
\end{equation}
directly from the tabulated equation of state and evaluate it along the maximum mass profiles shown in Fig.~\ref{fig:lqg_gamma_eta1}. The central values remain close to the relativistic degenerate limit and vary only mildly with $A_\lambda$. The main change is the radial interval over which the same thermodynamic response is sampled, because the stellar radius changes with the effective geometry. Thus the increase in maximum mass is not accompanied by an artificial stiffening of the matter sector. The plotted profiles are restricted to the interior interval $\Gamma\leq1.72$, which removes the low pressure edge of the tabulated lattice derivative and does not affect the central behavior used in the discussion.

Table~\ref{tab:selected_A_results} quantifies the displacement at the maximum of each equilibrium sequence. The first nonzero value, $A_\lambda=10$, already raises the maximum mass relative to GR while keeping the configurations inside the accepted numerical domain. For $A_\lambda=50$ and $A_\lambda=100$, the sequences reach still higher masses, with values in the super Chandrasekhar range for both equations of state. These configurations are calculated within the matter domain selected for the scan, including central densities below the inverse beta decay threshold of the equations of state, and therefore provide the relevant candidates for interpreting super Chandrasekhar configurations in the effective model.

\begin{table*}
\caption{Selected white dwarf configurations for the symmetric LQG case, $\eta=1$. The last two columns give the maximum absolute percent variation over $\eta=0.5$ and $\eta=2$ relative to $\eta=1$.}
\label{tab:selected_A_results}
\begin{ruledtabular}
\begin{tabular}{lccccc}
EOS & $A_\lambda$ & $M_{\max}$ [$M_\odot$] & $R_{\max}$ [km] & $|\delta M_{\max}|_{\eta}$ [\%] & $|\delta R_{\max}|_{\eta}$ [\%] \\
\hline
Chandrasekhar & 0 & 1.4166 & 1010.7 & 0.000 & 0.000 \\
Chandrasekhar & 10 & 1.4895 & 884.9 & 0.012 & 0.216 \\
Chandrasekhar & 50 & 1.6906 & 1495.9 & 0.081 & 0.326 \\
Chandrasekhar & 100 & 1.7125 & 2452.9 & 0.100 & 0.287 \\
lattice & 0 & 1.3850 & 980.5 & 0.000 & 0.000 \\
lattice & 10 & 1.4569 & 856.9 & 0.013 & 0.231 \\
lattice & 50 & 1.6586 & 1446.3 & 0.082 & 0.325 \\
lattice & 100 & 1.6907 & 2356.7 & 0.088 & 0.242 \\
\end{tabular}
\end{ruledtabular}
\end{table*}

The same table also shows why $\eta$ has little effect on the global observables in this regime. Changing the ratio from $\eta=1$ to $\eta=0.5$ or $\eta=2$ modifies $M_{\max}$ by less than $0.1\%$ and $R(M_{\max})$ by less than $0.33\%$ for the selected cases. This is smaller than the displacement produced by $A_\lambda$ and smaller than the separation between the two equations of state in the GR baseline. Within the present scan, the mass increase is therefore controlled by the polymer amplitude $A_\lambda$, while $\eta$ behaves mainly as a geometric asymmetry parameter.

\begin{figure}[H]
\centering
\includegraphics[width=\columnwidth]{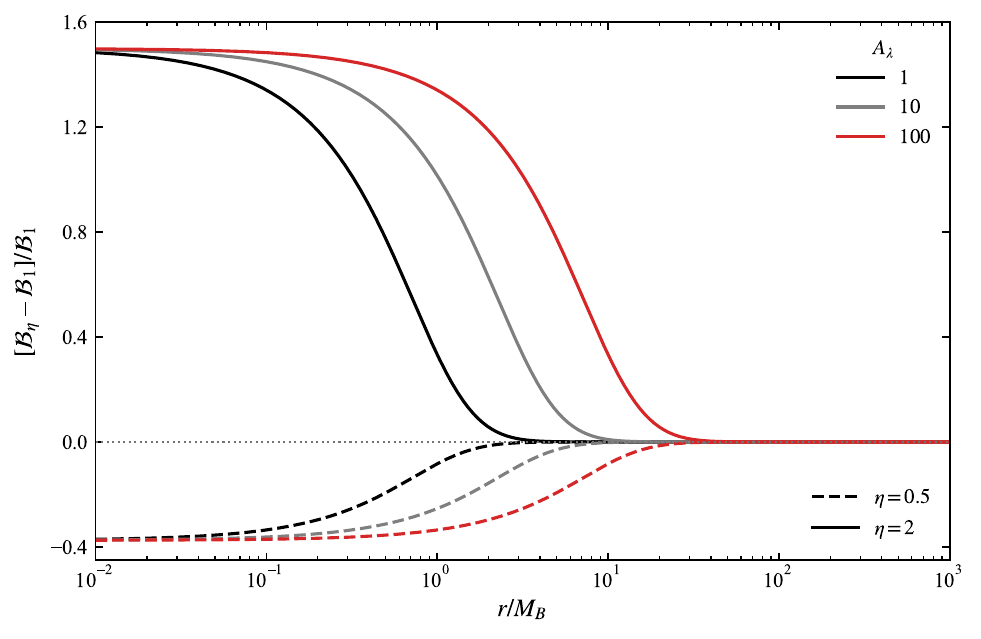}
\caption{Relative change of the metric function $\mathcal{B}(r)$ induced by $\eta=M_W/M_B$. The plotted quantity is $(\mathcal{B}_\eta-\mathcal{B}_{\eta=1})/\mathcal{B}_{\eta=1}$ as a function of $r/M_B$. Colors indicate $A_\lambda$, and line styles indicate $\eta$. The curves show that $\eta$ changes the metric near the transition region, while the relative effect decreases toward the exterior region.}
\label{fig:metric_B_eta}
\end{figure}

The weak response of the stellar observables to $\eta$ does not mean that $M_W/M_B$ drops out of the geometry. Figure~\ref{fig:metric_B_eta} shows that the asymmetry changes the metric function $\mathcal{B}(r)$ most strongly near the transition region of the polymer geometry. Away from that region, the relative difference decreases toward the exterior region. White dwarf configurations sample this weak field part of the solution rather than the transition surface itself. This separation explains why $\eta$ appears in the metric comparison but is nearly degenerate in the mass and radius sequences.

\section{Final remarks and perspectives}\label{con}

We have constructed and numerically implemented an effective Tolman Oppenheimer Volkoff system for carbon white dwarf configurations in the areal radius form of the polymer geometry. The prescription keeps the two mass parameters of the metric explicit through $M_B\rightarrow m(R)$ and $M_W=\eta m(R)$, so that the polymer amplitude $A_\lambda$ and the asymmetric ratio $\eta=M_W/M_B$ can be examined separately. This distinction matters because the two parameters do not play the same physical role in the stellar solutions.

The calculation first recovers the expected general relativistic behavior for the two carbon equations of state used here. The Chandrasekhar sequence and the sequence with the lattice correction define the reference scale for the mass and radius of the undeformed models. The lattice term lowers the maximum mass slightly, as expected from the reduced pressure support at high central density, but it does not change the qualitative structure of the standard sequence. This comparison fixes the scale against which the polymer correction is evaluated.

The main result is that $A_\lambda$ can move the equilibrium sequence into the super Chandrasekhar regime without changing the matter equation of state. In the accepted part of the scan, larger values of $A_\lambda$ shift the most massive configurations to higher gravitational masses for both equations of state. The configurations selected in Table~\ref{tab:selected_A_results} remain within the matter domain used in the calculation, including central densities below the inverse beta decay threshold of the equations of state. Within this effective model, they are therefore the relevant candidates for interpreting super Chandrasekhar white dwarfs.

The physical origin of this mass increase is geometric. The matter model is fixed, and the pressure support is not artificially increased by hand. Instead, the effective structure equations modify the balance between the pressure gradient and the gravitational term. This is why the same qualitative displacement appears for the Chandrasekhar equation of state and for the equation of state with the lattice correction. The result should be read as a property of the effective gravitational sector, not as an adjustment of the carbon microphysics.

The parameter $\eta$ has a different status. Varying $\eta$ changes the metric function $\mathcal{B}(r)$ and therefore records the asymmetry between the black hole and white hole mass parameters in the underlying geometry. For white dwarfs, however, the stellar configurations probe a low compactness region of the solution. In this regime the effect of $\eta$ on $M_{\max}$ and $R(M_{\max})$ remains at the subpercent level in the selected configurations. Thus $M_W/M_B$ is not absent from the geometry, but its imprint is largely suppressed in the global mass and radius observables considered here.

These conclusions are limited by the scope of the present calculation. The reported configurations are equilibrium solutions of the effective TOV system. We have not computed radial oscillation modes, secular stability, or evolutionary formation channels. The metric domain filters and the inverse beta decay condition define the admissible region used in the numerical scan, but they do not replace a full stability analysis. The parameters $A_\lambda$ and $\eta$ are also treated phenomenologically and are not inferred from observations in this version.

Several extensions follow directly from these results. The first is to compute radial stability for the super Chandrasekhar candidates and determine which part of the sequence survives the usual dynamical criteria. A second step is to repeat the calculation for additional white dwarf compositions and equations of state, so that the geometric effect can be separated from composition dependent reaction thresholds. It will also be necessary to compute baryonic masses and binding energies before interpreting $M_B$, $M_W$, or $M_\star$ as rest mass observables.

The same effective system can later be used in more astrophysical settings. Rotation, magnetic stresses, and finite temperature corrections are natural additions for massive white dwarfs, provided their contributions are introduced in the structure equations rather than absorbed into a parameter fit. On the observational side, the model can be compared with candidate super Chandrasekhar systems only after a likelihood and an uncertainty model are specified. The separation between the metric effect of $\eta$ and the sequence displacement produced by $A_\lambda$ also gives a useful guide for applying the construction to other compact objects, including neutron star models where the compactness is larger and the $\eta$ dependence may be less suppressed.

\begin{acknowledgments}
EO thanks the Funda\c{c}\~ao Cearense de Apoio ao Desenvolvimento Cient\'ifico e Tecnol\'ogico (FUNCAP), through grant BP6-0241-00335.01.00/25. MCA would like to thank FUNCAP, Funda\c{c}\~ao Cearense de Apoio ao Desenvolvimento Cient\'ifico e Tecnol\'ogico (Process No. DC3-0235-00076.01.00/24), and CNPq, Conselho Nacional de Desenvolvimento Cient\'ifico e Tecnol\'ogico, Brasil (Process No. 304145/2025-4), for financial support. JF would like to thank Funda\c{c}\~ao Cearense de Apoio ao Desenvolvimento Cient\'ifico e Tecnol\'ogico (FUNCAP) under grant PRONEM PNE0112-00085.01.00/16 and the Conselho Nacional de Desenvolvimento Cient\'ifico e Tecnol\'ogico (CNPq) under grant 304485/2023-3.
\end{acknowledgments}


\bibliographystyle{apsrev}
\bibliography{references}

\end{document}